\documentclass{emulateapj}
\usepackage{times}
\usepackage{graphicx}

\slugcomment{To Appear in ApJ}

\shorttitle{PSR J1741$-$2054's Bow Shock Nebula}
\shortauthors{}

\begin{document}

\title{The Balmer-dominated Bow Shock and Wind Nebula Structure 
of $\gamma$-ray Pulsar PSR J1741$-$2054}

\author{Roger W. Romani\altaffilmark{1}, Michael S. Shaw\altaffilmark{1}, Fernando Camilo\altaffilmark{2}, Garret Cotter\altaffilmark{3} \& Gregory R. Sivakoff\altaffilmark{4}}
\altaffiltext{1}{Department of Physics, Stanford University, Stanford, CA 94305}
\altaffiltext{2}{Columbia Astrophysics Laboratory, Columbia University, New York, NY 10027, USA}
\altaffiltext{3}{Department of Astrophysics, University of Oxford, Oxford OX1 3RH, UK}
\altaffiltext{4}{Department of Astronomy, University of Virginia, Charlottesville, VA 22904}
\email{rwr@astro.stanford.edu, msshaw@stanford.edu }

\begin{abstract}

	We have detected an H$\alpha$ bow shock nebula around PSR J1741$-$2054, 
a pulsar discovered through its GeV $\gamma$-ray pulsations. The pulsar is only
$\sim 1.5^{\prime\prime}$ behind the leading edge of the shock.  Optical spectroscopy 
shows that the nebula is non-radiative, dominated by Balmer emission.  The H$\alpha$ 
images and spectra suggest that the pulsar wind momentum is equatorially concentrated 
and implies a pulsar space velocity $\approx 150 {\rm \, km\,s}^{-1}$, directed 
$15\pm10^\circ$ out of the plane of the sky. The complex H$\alpha$ profile
indicates that different portions of the post-shock flow dominate line emission
as gas moves along the nebula and provide an opportunity to study the structure
of this unusual slow non-radiative shock under a variety of conditions.
{\it CXO} ACIS observations
reveal an X-ray PWN within this nebula, with a compact $\sim 2.5^{\prime\prime}$ 
equatorial structure and a trail extending several arcmin behind. Together these data support 
a close ($\le 0.5\;$kpc) distance, a spin geometry viewed edge-on and highly 
efficient $\gamma$-ray production for this unusual, energetic pulsar.
\end{abstract}


\section{Introduction}

	PSR J1741$-$2054, discovered in a blind pulsation search of a {\it Fermi}
point source \citep{blind}, is a $P=413$\,ms, characteristic age $\tau_c=391$\,kyr 
$\gamma$-ray pulsar. The pulsar was then detected in archival Parkes
radio observations and subsequently studied at GBT \citep{cet09},
showing it to be very faint and deeply scintillating.
The radio observations also gave a remarkably low dispersion measure
$DM= 4.7\,{\rm pc\, cm^{-3}}$, which for a standard Galactic electron density
model \citep{cl02} implies  a distance of $\approx 0.38$\,kpc. This makes this
one of the closest energetic pulsars known. At this distance the low
flux density observed at pulsar discovery, $S_{1.4} \approx 160 \mu$Jy, also makes this
the least luminous radio pulsar known. Indeed, since the source scintillates to
undetectability at many epochs, the time average radio luminosity is 
$L_{1.4}< 0.025 d^2_{0.4} {\rm \, mJy \, kpc}^2$.  The low distance was already
suspected from the relatively high $\gamma$-ray flux (for the modest
${\dot E} = 9.4 \times 10^{33} {\rm erg\, s^{-1}}$ spin-down luminosity).
At the $DM$-estimated distance, the pulsar would emit 28\% of its spin-down
power in $\gamma$-rays, if the radiation were isotropic.

	Since PSR J1741$-$2054 is an important addition to the set of nearby, energetic young
pulsars, further investigation of its energy loss and environs are needed.
In particular, we wish to constrain its motion, true age, non-photon
spindown deposition and interaction with its environment. For example,
it may be a significant source of $e^\pm$ injection in the solar neighborhood.
Such injection is generally visible as a pulsar wind nebula (PWN). The shape
of this nebula can constrain the source's distance, ISM interaction, and proper
motion. If the spin orientation can be measured \citep{nr08}, then we can also
model the $\gamma$-ray beam shape and connect the observed $\gamma$-ray flux 
to its true luminosity, predicting $f_\Omega = L_\gamma /(4\pi F_{obs} D^2)$
\citep{wet09}. This, in turn, provides the $\gamma$-ray efficiency, a critical
test of pulsar models.  

	We accordingly searched for a PWN, discovering an H$\alpha$ bow shock
and an X-ray PWN and trail. Like the few other pulsar bow shock nebulae, this
is a non-radiative, Balmer-dominated shock. The shock velocity is relatively low,
and Doppler broadened post-shock emission traces details of its flow,
presenting an opportunity for the study of collisionless shock structure under
unusual conditions, as well as constraining properties of the pulsar wind.
We describe initial observations and 
conclusions here. Additional discussion of the X-ray PWN and of the spectral
properties of the pulsar point source are presented in \citet{set11}. 

\begin{figure}[h!]
\vskip 6.0truecm
\includegraphics{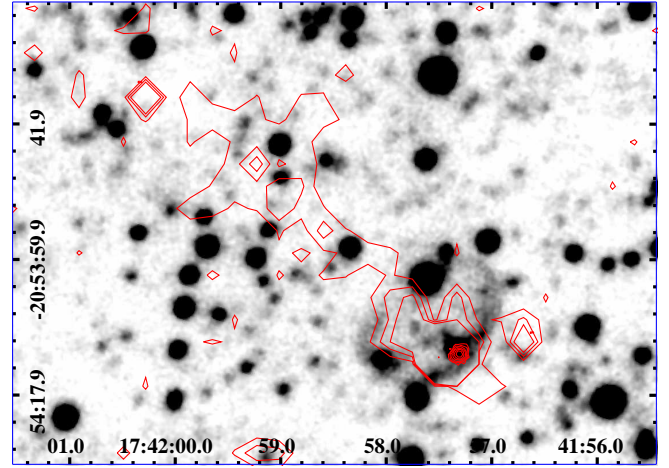}
\begin{center}
\caption{\label{WIYNim} WIYN 3.6m/MiniMo H$\alpha$ image of PSR J1741$-$2054's
bow shock.
Contours show the CXO 0.7-7\,keV emission, smoothed on a 14$^{\prime\prime}$
scale. Additional high contours smoothed at the 
1$^{\prime\prime}$ scale mark the PSR/PWN core.
}
\vskip -1truecm
\end{center}
\end{figure}

\section{Observations}

\subsection{H$\alpha$ Imaging}

	Our initial detection of the PSR J1741$-$2054 PWN was made in 
a 600s exposure using the WIYN 3.6m telescope and the MiniMo
camera with the KP W012 ($\lambda= 6566\AA,\,\Delta \lambda_{FWHM}= 16\AA$)
narrow-band H$\alpha$ filter on March 24, 2009. An additional 4$\times 300$s 
exposure plus a continuum frame
were obtained the next night. Unfortunately all data suffered from 
poor (1.4$-$1.8$^{\prime\prime}$) seeing. Nevertheless, these data revealed a clear elliptical
($13^{\prime\prime} \times 20^{\prime\prime}$) edge-brightened shell.
As this is close to the Galactic bulge ($l=6.4^\circ\, b=4.9^\circ$),
the continuum frame was crowded with field stars. Figure 1 shows the
H$\alpha$ image, without continuum subtraction. We were able to use the 
spectroscopic flux calibration (see below) to estimate 
the total (continuum-subtracted) H$\alpha$
flux of the nebula as $F_{{\rm H}\alpha} \approx 1.5 \times 10^{-14} {\rm erg\,
cm^{-2}s^{-1}}$.

\begin{figure}[h!]
\vskip 5.9truecm
\includegraphics{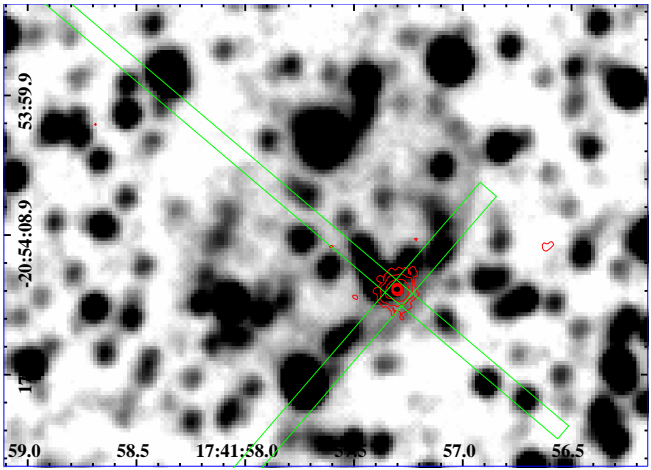}
\begin{center}
\caption{\label{NTTim} NTT 3.6m/EFOSC2 H$\alpha$ image of the PSR J1741$-$2054
bow shock.
Contours show the CXO 0.7-7\,keV emission, smoothed on a 0.4$^{\prime\prime}$
scale. The two rectangular boxes show the 1$^{\prime\prime}$ wide long-slit 
positions of the Keck LRIS spectroscopy.
}
\vskip -1truecm
\end{center}
\end{figure}

	We subsequently (Aug 21, 2009) obtained 3$\times$600\,s exposure
using the EFOSC2 camera on the NTT 3.6m through the H$\alpha$\#692 filter.
While the image quality was somewhat better ($1.1^{\prime\prime}$), the
wider ($\Delta \lambda_{FWHM} = 62\AA$) transmission band and lower 
peak throughput gave the H$\alpha$ nebula sharper edge structure, but 
lower contrast (Figure \ref{NTTim}). On this image we
show the position of the {\it CXO} point source and the locations
of the $1^{\prime\prime}$ slits used for the spectroscopic study.

\subsection{X-ray Imaging}

	We obtained a {\it CXO}/ACIS-S observation (ObsID 11251) of the pulsar 
field on May 21, 2010 (=MJD 55337.1), with 48.8\,ks of exposure. The pulsar
was placed near optimum focus on the backside illuminated S3 chip
and the CCDs were operated in half-frame VFAINT mode, reading out every 
1.64\,s to reduce pileup.
A bright point source and compact nebula are seen, locating the pulsar to 
17h 41m 57.28s, $-$20$^\circ$ 54$^{\prime}$ 11.8$^{\prime\prime}$ 
($\pm 0.3^{\prime\prime}$). 
The crowded optical field made it difficult to identify field X-ray sources.
However, over the S3 half-chip several soft (coronal emission) sources were 
clearly identified with field stars, confirming the X-ray/optical frame tie
at the $\sim 0.3^{\prime\prime}$ level. 

There is diffuse, faint X-ray emission in a PWN trail extending some
$2^\prime$ at position angle PA=$45\pm5^\circ$ (N through E). This 
structure contains $\sim 900\pm 100$ counts. From 2$-7^{\prime\prime}$
from the pulsar the trail contains an additional 92$\pm14$ counts. Finally,
the compact core contains $\sim 3460$ counts in a $2^{\prime\prime}$ 
radius region about the pulsar, dominated by the point source. 

	The emission from these structures is consistent with a simple
absorbed power law with spectral index $\Gamma \approx 1.6\pm 0.2$ and
absorbing column $N_H \approx 1.5 \pm 0.3 \times 10^{21} {\rm cm^{-2}}$.
The point source clearly shows an additional soft thermal component, and the
detailed spectral measurements of these structures are discussed in
\citet{set11}. Here we concentrate on the PWN morphology.
\medskip

\begin{figure}[h!]
\vskip 5.5truecm
\includegraphics{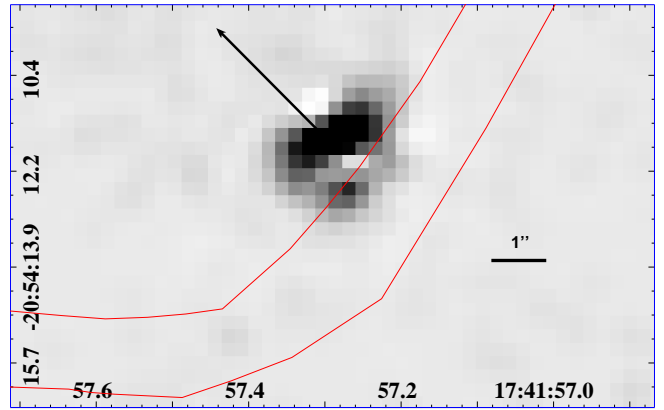}
\begin{center}
\caption{\label{PSFsub} PSF-subtracted PWN core, showing
extended emission (excess in black). If interpreted as an
equatorial torus, the symmetry axis lies within 5$^\circ$ of the large scale
trail axis (arrow) and the H$\alpha$ bow shock symmetry axis. The approximate
location of the H$\alpha$ limb is shown by the curved band.
X-ray photons ($0.7-7\;$keV) were binned to 0.25$^{\prime\prime}$ pixels
and smoothed with a $\sigma=0.5^{\prime\prime}$ Gaussian after subtraction of a
scaled MARX PSF.
}
\vskip -0.5truecm
\end{center}
\end{figure}

	The region around the point source appears slightly extended.
After subtracting a PSF generated with MARX, shifted and scaled to
best match the point source component, the excess appears dominated by
a diagonal band, $\approx 0.75^{\prime\prime} \times 2.5^{\prime\prime}$,
with a minor axis at PA=$40\pm5^\circ$. This structure appears to contain
$\sim 400\pm 50$ counts or 11\% of the total counts in this region.
As the minor axis PA is within 5$^\circ$ of the large scale X-ray trail
axis (Figure \ref{PSFsub}), we associate this structure with the equatorial 
torus of the PWN.
The 3:1 aspect ratio suggests a planar structure seen edge on, and
implies that the pulsar spin axis is close to the plane of the sky.

\subsection{Nebular Spectroscopy}

	Pulsar H$\alpha$ bow shock nebulae are relatively rare, since
the pulsar wind must impact a partly neutral medium. 
Typically, this means that the pulsar has escaped its supernova remnant 
birthsite and is interacting with the general ISM \citep{cc02}. The
shock spectra, however, are particularly interesting as the shocked
gas is advected downstream  on a time short compared to the cooling
time. This means that these are non-radiative shocks and are dominated
by Balmer emission rather than the usual forbidden lines \citep{ray01,h10}.
When the up-stream neutral atoms, which are unaffected by the 
fields in the shock, drift into the shocked ISM gas they can be
excited, producing line radiation at the (cold) radial velocity
of the ambient medium gas. For typical conditions the ratio of
excitation to ionization rates is $q_{ex}/q_i \approx 0.25$ so that
we obtain an ISM-velocity Balmer photon for every $\sim 4$ neutral atoms. 
However, in addition, the atoms can charge-exchange with the hot shocked
ions. This gives rise to excited neutrals in the post-shock flow and 
a broad line component at the systemic radial
velocity of the post-shock gas. Thus spatially-resolved,
kinematically-resolved observations of the line emission can provide
a wealth of information on the local ISM and bow-shock kinematics
\citep{arc02}.

	We were able to obtain two $1^{\prime\prime}$ long-slit 
spectra of the PSR J1741$-$2054 bow shock, using the Keck I/LRIS on June ll
2010 (MJD 55358.5). A total of 900s exposure was taken with the slit at
${\rm PA=} 50^\circ$ (along the PWN symmetry axis) and 1200s total exposure
with the slit at ${\rm PA=} 140^\circ$.
Both exposures covered the pulsar position. Unfortunately, since only
one amplifier was available in the LRIS red camera, the long slit
was truncated. However, we managed to cover the PWN emission; the slit
locations were accurately determined from recorded slit images and are
shown on Figure 2.

\begin{figure}[h!]
\vskip 7.2truecm
\includegraphics{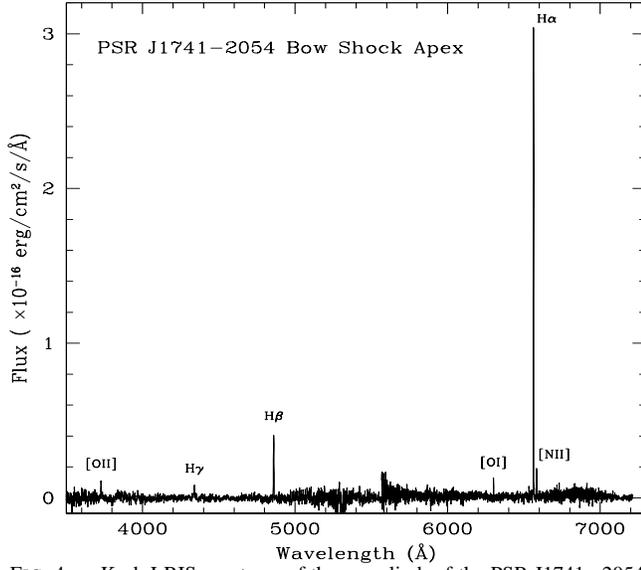}
\begin{center}
\caption{\label{BSspec} Keck LRIS spectrum of the apex limb of the 
PSR J1741$-$2054 bow shock. The extreme Balmer dominance is evident,
although the strongest forbidden lines are weakly detected from the
bright patches on the limb.
}
\end{center}
\end{figure}

	For these observations, a 5600\AA\, dichroic separated the red
and blue beams. The blue camera employed a 600l/mm, 4000\AA\, blaze grating
and covered 3100-5600$\AA\,$ at a resolution of $\Delta \lambda = 0.63\AA$/pixel.
For the bow shock exposures, the red camera employed the 1200l/mm, 7500\AA\, 
grating for a resolution of $\Delta \lambda = 0.4\AA$/pixel. With the
1$^{\prime\prime}$ slit the achieved resolution as measured from the night
sky lines near H$\alpha$ was 1.48\AA\, or 68${\rm \, km\,s}^{-1}$ (FWHM).
The spatial resolution along the slit was 
0.24$^{\prime\prime}$/pixel. 
For the blue channel, we were able to employ standard calibrations. 
Unfortunately the red channel configuration was non-standard for
the observing run and a flux standard was not obtained in this configuration.
However, we were able to use flux measurements at a lower resolution setting
along with measured grating responses to recover the red flux calibration
with a conservative estimate of 15\% accuracy. The wavelength calibration
was refined by reference to the night sky lines and line velocities have
been converted to the Local Standard of Rest (LSR) using the \citet{mb81}
constants.

Figure \ref{BSspec} shows the full spectrum of the bright limb in front of the
pulsar from the $PA = 50^\circ$ observation. The strong dominance of 
the Balmer emission is clear, although a few of the brightest forbidden
lines are also visible. These lines were detectable only in the brightest 
portions of the shock limb, so we have not been able trace this component
separately.  Table 1 contains measurements of the line fluxes.

\begin{deluxetable}{lcrl}[h!!]
\tablecaption{\label{Lines} PSR J1741$-$2054 Bow Shock Limb Fluxes}
\tablehead{
  \colhead{Species} & \colhead{Wavelenth} & \colhead{Flux} &\cr 
   & $\AA$& $10^{-17}{\rm erg\,cm^2 s^{-1}}$ &
}
\startdata
H$\alpha$ & 6563 & 70.2$\pm 10.5$& $^a$ \\
H$\beta$  & 4861 & 17.6$\pm 0.5$&  \\
H$\gamma$ & 4340 & 5.2$\pm 0.6$&  \\
H$\delta$ & 4102 & 1.3$\pm 0.5$&  \\
N[II]     & 6583 & 4.0$\pm 0.7$& $^a$ \\
N[II]     & 6549 & 1.0$\pm 0.4$&  \\
O[I]      & 6300 & 1.6$\pm 0.3$&  \\
O[I]      & 6363 & 0.3$\pm 0.2$&  \\
O[II]     & 3727 & 4.8$\pm 0.5$&  \\
S[II]     & 6731 & 1.0$\pm 0.2$&  \\
S[II]     & 6716 & 1.2$\pm 0.3$&  \\
\enddata
\tablenotetext{a}{Error dominated by calibration systematics}
\end{deluxetable}
\bigskip

    Our Balmer measurements give H$\alpha$/H$\beta = 3.99\pm 0.61$
and H$\gamma$/H$\beta = 0.295\pm 0.035$, and we can use these flux ratios
to estimate the extinction toward the nebula. The selective extinction is
$$
E(B-V) = 2.21{\rm Log}({{\alpha/\beta}\over{\alpha/\beta_m}}),\,
= -5.17{\rm Log}({{\gamma/\beta}\over{\gamma/\beta_m}})
$$
where the models values (e.g. $\alpha/\beta_m$) depend on the shock populations
and geometry, especially the shock optical depth of Ly$\beta$.
\citet{ckr80} have estimated the Balmer ratios for the broad
component of a radiative shock (their Model 2). For the optically
thin (Case A) situation, they find $\alpha/\beta_m=2.9$ and
$\gamma/\beta_m=0.39$. Our corresponding extinction estimates
are $E(B-V)=0.31\pm0.15$ and $0.63\pm0.26$; the $1\sigma$ overlap range 
is $E(B-V)=0.37-$0.46. However if the Ly$\beta$ optical depth
is $\sim 3-5$, trapping increases the expected H$\alpha$ flux and
the predicted broad line ratios are $\alpha/\beta_m=4.8$ and
$\gamma/\beta_m=0.32$. The H$\alpha$ values then gives $E(B-V)<$0;
physical, positive values are allowed only at the $\sim 1.3\sigma$ level. 
The
H$\gamma$ flux gives $E(B-V)=0.18\pm0.25$. Thus we conclude that the
selective extinction is $< 0.45$ with smaller values preferred for 
finite thickness shocks, especially for the H$\alpha$ measurement.
For reference, the fitting formulae for extinction values in \citet{cet98} 
give $E(B-V)=A_V/3.1=0.13$ for this direction and $d=0.4\,$kpc.
Our limit on the extinction can be translated to an effective 
Hydrogen column $N_H \approx 5.6 \times 10^{21}  E(B-V) {\rm cm^{-2}} <
2.5 \times 10^{21} {\rm cm^{-2}}$, in agreement with the
X-ray spectral fitting of the PWN. For either our conservative
upper limit or the lower extinction inferred for modest optical
depth shocks the neutral column is substantially larger than the
low $N_e \approx 1.5 \times 10^{19} {\rm cm^{-2}}$ inferred from the
DM, suggesting low ionization along this line of sight.

\begin{figure}[b!!]
\vskip 7.7truecm
\includegraphics{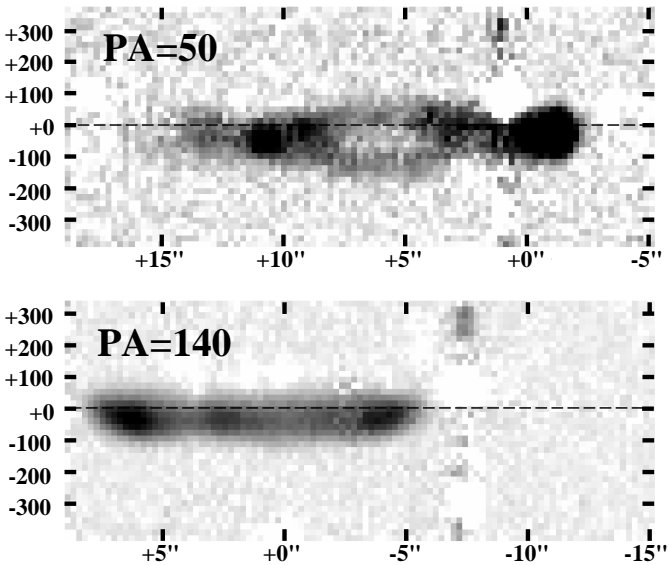}
\begin{center}
\caption{\label{SlitSpec} Symmetry- (PA=$50^\circ$, above) and cross- 
(PA=140$^\circ$, below) axis long slit spectra of the PSR J1741$-$2054 
bow shock. Both pass through the estimated pulsar position at coordinate 
0$^{\prime\prime}$. The spectra are sky- and continuum-subtracted and 
velocity shifts (${\rm km\,s^{-1}}$) are with respect to the LSR; residuals
from bright stars appear, particularly 1.5$^{\prime\prime}$ behind the pulsar
(PA=$50^\circ$) and 7$^{\prime\prime}$ to the right of the pulsar
(PA=140$^\circ$).  The symmetry-axis spectrum shows red- and blue-shifted
components from the front and back of the PWN shell. The cross-axis spectrum
is dominated by blue-shifted emission near the front limb of the PWN. 
}
\vskip -0.5truecm
\end{center}
\end{figure}

	With the spectral resolution and S/N available we were able to
resolve the H$\alpha$ lines, making an initial exploration of the velocity
structure in the post-shock flow. In Figure \ref{SlitSpec} we show the
H$\alpha$ spectra along the two slit axes, with velocity shifts measured
relative to the LSR as determined from night sky lines.
Emission from the approaching (blue) and receding (red) sides
of the bow shock are clearly seen in the PA=$50^\circ$ panel. 
Interestingly the front of the bow shock appears to be dominated by
blue-shifted emission. We discuss a likely interpretation of this peculiar
velocity structure below.

\section{Bow Shock Modeling}

	The spatial and kinematic information provided by these observations
give an excellent opportunity to probe the geometry of the PWN outflow
\citep{arc02}. Very helpful in this study are the elegant solutions to the
thin momentum-conserving bow shock provided by \citet{wil96,wil00}. Although
these analytic results strictly only describe the shape of a contact discontinuity
for a thin, fully mixed cold flow, they do provide a good approximate shape
for the non-radiative forward shock in the ISM \citep{ck98,b02}. The general
velocity structure also follows the numerical simulations, although the cold
analytic model has somewhat denser flows and correspondingly higher velocities 
downstream from the bow shock apex.
The characteristic standoff angle of the contact discontinuity in the thin
shock limit is
$$
\theta_0 = R_0/d = \left ( {{{\dot m_W}v_w} \over {4\pi\rho_0 v_{\rm psr}^2}} \right )^{1/2}/d
= 2.1^{\prime\prime}{\dot E}_{34}/(n\, v_7^2\,d_{0.4})^{1/2}
$$
where for a relativistic pulsar wind the momentum flux is ${\dot m_W}v_w = {\dot E}/c$
and the numerical value is scaled to the nominal spindown flux and distance of PSR
J1741$-$2054 traveling at 100$v_7{\rm \, km\,s}^{-1}$ through a medium of number density $n$.
Thus we expect a small stand-off angle for the bow shock, as seen in Figure 1. 
Numerical simulations show that the forward ISM shock scale is a factor 
$\sim 1.3\times$ larger \citep{bb01}.

	However, Figure 1 shows that the observed H$\alpha$ shock is very flat
at the apex; it does not fit well to the standard \citet{wil96} thin shock solution.
While the numerical thick bow shock models do show a somewhat wider angle
spread downstream, they do not display the very small standoff and flattened
apex of the PSR J1741$-$2054 bow shock. Effects that may produce such 
flattened structure are underlying wind asymmetry and projection to the Earth line-of-sight.
\citet{wil00} has provided useful expressions for the structure of anisotropic
winds. In the pulsar case, we expect that the wind momentum flux will be axisymmetric
about the pulsar spin axis. In general this provides momentum deposition misaligned
with the bow shock axis. However, there is good evidence \citep{jet05,nr08} that pulsar
spin and proper motion axis have a tendency to be aligned. We therefore focus
on the aligned, axisymmetric wind.

	The pulsar wind is also highly relativistic, so that the mass of the
bow shock shell is dominated by the swept-up mass of the ISM. In the nomenclature
of \citet{wil00} this is $\alpha=v_{PSR}/v_w \longrightarrow 0$, while the aligned
case is $\lambda=0$. We further restrict to an equatorially symmetric wind
($p_w=m_w v_w \propto c_0 + c_2 {\rm cos}^2\theta$). An isotropic wind has $c_2=0$
while an equatorial ($p_w \propto {\rm sin}^2\theta$) wind has $c_2=-3/2$. For
these conditions the analytic solution is a surface of rotation with pulsar
shock distance $R(\theta$) for $\theta$ the angle to the direction of the pulsar-ISM
velocity with a cylindrical coordinate
$$
{\tilde {\bar \omega}}^2= (R/R_0)^2{\rm sin}^2\theta 
= 3( 1 - \theta\;{\rm cot}\theta)(1-c_2/12) + 3 c_2{\rm sin}^2\theta\;/4.
$$
We can also write the tangential velocity (in the bow shock frame) and surface density 
of  swept-up mass in the cold shell in terms of two quantities 
$$
G_{\tilde{\bar \omega}}=[(1-c_2/12)(2\theta-{\rm sin}2\theta)+{\rm sin}^3\theta\;{\rm cos}\theta]/4
$$
and
$$
G_z=[(1-c_2/12)(1-{\rm cos}2\theta)+c_2/2 \, (2+{\rm cos}2\theta){\rm sin}^2\theta]/4
$$
with
$$
v_t = v_{PSR} [4 G_{\tilde{\bar \omega}}^2 + (2 G_z - {\tilde {\bar \omega}}^2)^2]^{1/2}/{\tilde {\bar \omega}}^2
$$
the tangential surface velocity in the shell,
$$
\sigma = n\; \mu m_p \; R_0 ({\tilde {\bar \omega}}^2)^{1/2} v_{PSR}/(2 v_t)
$$
the shell surface mass density, and $\mu m_p$ the mean mass per particle of the ambient ISM.

	This bow shock structure will be axisymmetric in some angle $\phi$
about the pulsar velocity, which will be inclined by angle $i$ to the Earth 
line-of-sight. To complete the model of the bow shock appearance and radial 
velocity structure we project to the plane of the sky.
This places the emission at projected angles $(\zeta, \chi)$ with respect to
the (pulsar) wind origin with 
$$
{\rm cos} \zeta = {\rm cos}\theta \;{\rm cos} i + {\rm sin}\theta \;{\rm sin} i \;{\rm cos}\phi
$$
and 
$$
{\rm sin}\chi = {\rm sin}\theta \;{\rm sin}\phi / {\rm sin} \zeta.
$$
The emission from each point along the bow shock is proportional to the 
appropriate $\sigma (\theta, \phi)$ and the projected velocity (in the observer frame) is 
given by 
$$
v_{bs}=v_t ( {\rm cos} \gamma \;{\rm cos} i + {\rm sin}\gamma \;{\rm sin} i \;{\rm cos}\phi) 
- v_{\rm psr}\;{\rm cos} i
$$
with $\gamma=\theta + {\rm tan}^{-1}[R/(dR/d\theta)]$.
\smallskip

	Note that this velocity solution is for the mixed (cooled, zero pressure)
flow of the swept up gas. In practice, numerical simulations show
modest departures from this analytic solution for the tangential velocity 
\citep{b02} and a detailed solution for the full line structure of the nebula
should consider such effects. However, a more basic difference lies in the fact that
the non-radiative H$\alpha$ arises from neutrals drifting into, and charge exchanging
with, ions in the post-shock flow. Thus the H$\alpha$ structure depends strongly
on the portion of this flow which is reached by such neutral atoms \citep{bb01}.
In particular, immediately behind the forward (ISM) shock we expect that
the gas has velocity $3/4 v_{\rm psr} {\rm cos} \eta$, where the angle between 
the pulsar velocity and the shock normal is $\eta = \theta - {\rm tan}^{-1}[(dR/d\theta)/R]$.
Only later does the post-shock flow converge to the tangent to the contact
discontinuity. This effect may be seen in the simulations of \citet{ck98}.
Thus immediately behind the ISM shock one expects a line-of-sight velocity
$$
v_{nr} = -3/4 v_{\rm psr} {\rm cos} \eta \; {\rm cos} i.
$$
Since ${\rm cos} \eta$ is always positive, this velocity always
has the sign of $-{\rm cos} i$.

\subsection{Comparison with the Observations}

\begin{figure}[b!!]
\vskip 7.9truecm
\includegraphics{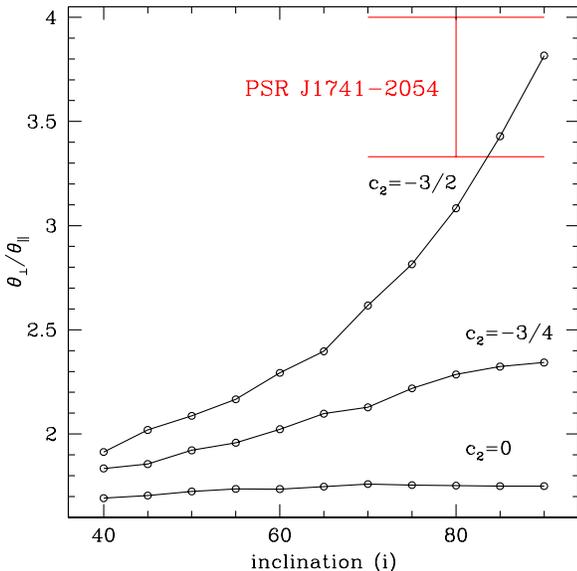}
\begin{center}
\caption{\label{Shockshape} Ratio of the perpendicular to standoff angular
sizes as a function of $i$, the inclination angle of the pulsar velocity.
The large  ratio observed for PSR J1741$-$2054 is shown, suggesting
an equatorially dominated wind and a pulsar velocity nearly in the plane of
the sky.
}
\end{center}
\end{figure}

	An initial question is whether the shock created by an 
equatorial wind can reproduce the flattened shape of this pulsar bow shock. One
way to parametrize this shape is the ratio of the perpendicular
half angular size $\theta_\perp$, measured through the pulsar, 
to the angle from the pulsar to the projected
limb of the wind shock in the forward direction $\theta_{||}$. Note that
$\theta_{||} \ne \theta_0 {\rm sin} i$.
This ratio is very large for PSR J1741$-$2054's bow shock, $\approx 3.7
\pm 0.3$.  We have computed this ratio for a number of axisymmetric, aligned wind bow
shock models.
Figure \ref{Shockshape} shows that the projected shock limb has a ratio nearly
independent of $i$ for an isotropic ($c_2=0$) wind, but that the ratio
can reach the observed value for a ${\rm sin}^2\theta$ ($c_2=-3/2$) wind
and a pulsar motion nearly in the plane of the sky. Considering winds that are 
even more equatorially concentrated or including the flaring effect
of the finite pressure in a thick bow shock allows somewhat smaller $i$ to 
be accommodated.

\begin{figure}[t!!]
\vskip 7.0truecm
\includegraphics{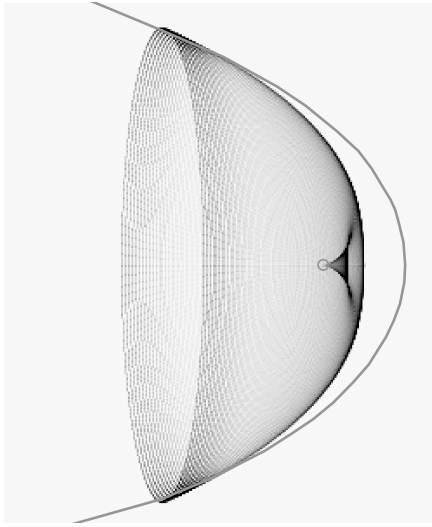}
\begin{center}
\caption{\label{Mod_im} Projected bow shock shape for an
equatorial ($c_2=-3/2$) wind with pulsar velocity inclination $i=80^\circ$.
The pulsar position is marked with a circle. Note that the projected bow
shock limb has an appreciably flatter apex than that of an
isotropic ($b_3=0$) edge-on ($i=90^\circ$) shock (line).
}
\end{center}
\end{figure}

	Figure \ref{Mod_im} shows the projected shape of an equatorial
relativistic wind emanating from the pulsar position (circle) for an inclination
$i = 80^\circ$. For comparison, the line shows the limb of the isotropic
wind bow shock; the standoff distance for the equatorial wind is 
appreciably reduced. Note that, since formally $p_w \propto {\rm sin}^2\theta$,
the bow shock approaches the star in the forward ($\theta=0$) direction. Since the pulsar wind
likely has a jet component, a physical bow shock would be smooth at the apex.
Similar matching to the H$\alpha$ limb was used by \citet{get02} to argue that
the wind of the millisecond pulsar PSR J2124$-$3358 is anisotropic. For
PSR J1741$-$2054 we have additionally been able to connect the shock shape
with anisotropy detected in the X-ray synchrotron nebula (\S 2.2).

\begin{figure}[b!!]
\vskip 3.9truecm
\includegraphics{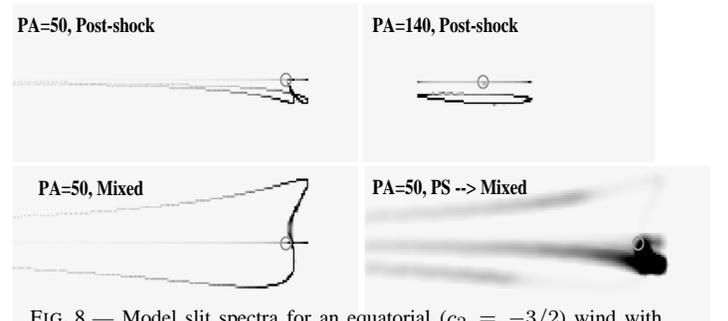}
\begin{center}
\caption{\label{Mod_spec} 
Model slit spectra for an equatorial ($c_2=-3/2$) wind with 
pulsar velocity inclination $i=70^\circ$. The slit axes are at $10^\circ$
to the pulsar velocity vector and the pulsar position is marked (circle) at the
systemic (narrow component) velocity. Above: prompt emission from the
symmetry and cross axes.
Below: The expected velocity structure of the `mixed' gas emission for
the PA=50$^\circ$ slit, and a smoothed model where the mixed gas 
acquires a neutral component, appearing downstream.
}
\end{center}
\end{figure}

	We next check if our model can explain the features of Figure \ref{SlitSpec}.
Most remarkable is the dominance of blue-shifted emission at and
in front of the pulsar. For the PA=$50^\circ$ slit, the bright knot at the
apex is at $v_r = -34$ to $-44{\rm \, km\, s}^{-1}$. The cross-axis slit has H$\alpha$ offset to
$v_r = -29$ to $-44{\rm \, km\,s}^{-1}$, with no equivalent red-shifted component. 
This is significantly offset from the systemic (ISM) velocity. 
The lack of any red-shifted emission is at first
surprising, given that mass flows around the bow shock. However, if the pulsar
is moving out of the plane of the sky ($i < 90^\circ$), we see that the prompt 
post-shock broad-line emission (from neutral atom charge exchange) will show
negative radial velocities from both
sides of the nebula. This is illustrated in the top two panels of Figure
\ref{Mod_spec}, which trace the velocity structure of the prompt emission.
Here we plot a model for $i=70^\circ$ to better separate the velocity
components. The $i=75-80^\circ$ models (for  a 3D space velocity
$v_{\rm psr} \approx 150\pm 50{\rm \, km\,s}^{-1}$) that match the shock shape also provide 
the best match to the velocity shifts in the slit spectra.

	The immediate post-shock layers can dominate if the neutrals
penetrate only partly into the shocked ISM, before suffering nearly
complete ionization. This is equivalent to Case C of \citet{bb01}.
In fact PSR J1741$-$2054 has a relatively large spindown luminosity among
pulsars showing ISM H$\alpha$ shocks (although below that of PSR B0740-28),
so if the density and ionization fraction of the upstream medium are high,
ionization may indeed be strong in the shocked ISM. Thus, ahead of the pulsar
where the shock is nearly normal, neutral H may only exist in the 
immediate post-shock layer and blue shifted emission would dominate the 
observed spectrum.

	The pulsar space velocity $v_{\rm psr}$ is modest and as one
moves behind the pulsar position, the shock obliquity $\eta$ increases
sharply and the post-shock heating drops. As ionization drops,
neutral H may reach the bulk flow along the contact discontinuity
(Bucciantini \& Bandiera 2001 Case B). Charge exchange will continue
and thus we expect components with negative radial velocity (near side) 
and positive radial velocity (far side).
Indeed in Figure \ref{SlitSpec}, across the body of the shell, the
PA=$50^\circ$ spectrum shows two components with velocity extrema 
$-125{\rm \, km\,s}^{-1}$ and $+25{\rm \, km\,s}^{-1}$ 
from the front and back sides of the flow. 
Figure \ref{Mod_spec} shows the computed radial
velocity for this `mixed component'. The large positive and
negative velocities expected at the pulsar position are not seen in the PA=$140^\circ$
spectrum or in the PA=$50^\circ$ spectrum in front of the pulsar, 
suggesting that near the apex neutral H atoms do not penetrate to this `mixed'
tangential portion of the flow. The lower right panel shows a heuristic merged 
model, where the `mixed' component grows as $\theta$ increases
downstream. This smoothed model bears reasonable similarity to the observed
2D spectrum, although it fails to reproduce the `closed off' 
shape along the symmetry axis and the red-shifted `mixed' emission from the body
of the nebula is less prominent in the data.  Clearly additional
effects are needed to reproduce the full velocity structure of this bow shock
and additional imaging and spectroscopy will be needed to fully explain its origin.
For example, the close-out of the symmetry axis spectrum suggests an expanding
bubble structure rather than an open-backed bow shock. Bubble structure has
been seen for other PWN bow shocks (e.g PSR B2224+65's `guitar nebula',
Cordes et al. 1993); \citet{vKI08} have proposed  that such bubble
structures are a natural consequence of instabilities in the 
post-shock flow of the relativistic pulsar wind. The fainter and slower red wing
`mixed' emission 5-10$^{\prime\prime}$ behind the pulsar might be due to
ISM inhomogeneities or, plausibly, instabilities in the PWN trail.

	One additional difference to the data is the lack of 
an obvious ISM-velocity `narrow' line component in the measured spectra.
In the model spectra, the component is assigned 10\% of the broad-line flux.
The limited LRIS spectral resolution makes it difficult to discern any
such component, particularly in the brighter knots of the bow shock emission.
However, we estimate from the PA=$50^\circ$ spectrum of the nebular body
a broad-to-narrow intensity ratio $I_b/I_n > 3$. Interestingly, this ratio provides a measure
of $T_e/T_p$, probing electron-ion equilibration in the post-shock emitting
region \citep{ray01}; relatively large ratios and fast equilibration appear
to occur in low velocity shocks \citep{vaet08}, supporting a modest velocity for PSR
J1741$-$2054.  If the narrow component can be isolated and 
kinematically resolved, it provides much additional information on the
upstream medium's density and on pre-shock heating of the
ambient gas.

\section{Geometry and Distance Constraints}

	The leading edge of the bow shock appears structured, especially 
in the NTT image, so it is somewhat difficult to measure the pulsar-apex angle
and cross-axis half angle.  Our best estimates are 
$\theta_{||} \approx 1.5^{\prime\prime}$ and 
$\theta_\perp \approx 5.5^{\prime\prime}$. 
We would like to relate these to the characteristic contact discontinuity
standoff angle $\theta_0$. In \citet{arc02}, the ISM shock/contact discontinuity
angle ratio was estimated as $\sim 1.3$ at the shock apex. While finite
pressure effects can cause this to grow slightly
downstream, here we assume a constant ratio. Also, for the 
$c_2=-3/2$, $i=80^\circ$ shape, we infer $\theta_{||} = 0.5 \theta_0$
and $\theta_\perp = 1.85 \theta_0$. Together these estimates let us infer
a characteristic (isotropic wind) contact discontinuity standoff angle 
$\theta_0 \approx 2.3^{\prime\prime}$.

	This characteristic offset implies $n\;d_{0.4}v_7^2 =0.8$. As noted above
the velocity structure suggests that $v_7 \approx 1.5\pm 0.5$, so
we loosely constrain the product $n\;d_{0.4} \approx 0.4$. In turn
this suggests a plausible density $n\approx 0.4 \;{\rm cm^{-3}}$ 
if we adopt the DM distance.

	For the allowed extinction range ($A(H\alpha)=0 - 1.1$; i.e 
$\tau_{\rm Ly\beta}=0-5$), we infer an H$\alpha$ number flux from the full nebula of 
${\dot N}_\gamma = 5-15 \times 10^{-3} {\rm cm^{-2}s^{-1}}$. 
This may be related to the neutral density and velocity of the upstream
medium \citep{ray01,get01}. For example the upstream neutrals will
produce a narrow component H$\alpha$ number flux
$$
{\dot N}_{\gamma, n}\approx {{q_{ex} n_{HI} \pi  (\theta_{eff} d)^2 v }
\over{q_i} 4\pi d^2 }=
1.5 \times 10^{-3} n_{HI}\theta_{10}^2 v_7 {\rm cm^{-2}s^{-1}}
$$
for an excitation-to-ionization ratio $q_{ex}/q_i=0.25$ and effective
full nebula angular radius $\theta_{eff} = 10^{\prime\prime} \theta_{10}$.
This predicts $\le 0.3$ of the observed flux of the (broad) H$\alpha$ 
line.  One can make a similar estimate for the flux of the broad component,
depending on the ratio of charge-transfer to ionization rates,
$q_{ct}/q_i$, i.e. $(q_{ct}/q_i) n_{HI} \theta_{10}^2 v_7 = 3-10$ (depending
on the extinction). With the low $v_7 \approx 1.5$ and $n \le 1$ estimated
above, we would infer a large yield of broad $H_\alpha$ photons from
charge transfer interactions, $q_{ct}/q_i > 2-7$. Figures in \citet{vaet08}
suggest that the broad to narrow ratio grows rapidly at low shock velocities,
especially for optically thin shocks, although velocities as small as 150 
${\rm km\,s}^{-1}$ are not computed. Additional modeling and, especially, an
accurate measurement of the narrow line component would help settle whether
such high efficiency conversion to H$\alpha$ occurs.

	Turning to the nebula distance, we see that the relatively low
$N_H$ inferred from the X-ray absorption column and the Balmer line ratios
supports a close distance for this pulsar, with the HI surveys \citep{dl90,ket05}
showing values twice that inferred for the nebula at distances as small as $0.5\;$kpc
({\it HEASARC nH tool}). With improved spectral measurements of the PWN 
H$\alpha$ we should be able to fit for a precise pulsar space velocity
and inclination.  Comparison with a proper motion, measurable from 
{\it CXO} X-ray imaging ($\sim$ 7y baseline) or possibly HST imaging
(few year baseline) would then yield a direct kinematic distance to the 
neutron star. This is particularly valuable for understanding the apparently
very high $\gamma$-ray efficiency.

	Our match to the $\theta_{||}/\theta_\perp$ ratio suggests 
$i\approx 75\pm 10^\circ$. For our estimated $v_p \approx 150\,{\rm km\, s^{-1}}$
this is also consistent with the spread of velocities in our slit spectra 
and with the measured apex blue shift 
$\sim -30 {\rm km \, s^{-1}} \approx (-3/4) {\rm cos}(75^\circ) 150 \,{\rm km\, s^{-1}}$
expected if the surrounding medium radial velocity is near the LSR and
prompt emission dominates near the apex.
If the pulsar spin and velocity are aligned,
this implies a pulsar viewing angle $\zeta \ge 65^\circ$. In turn this may be
compared with the viewing angles inferred for the observed $\gamma$-ray pulse
profile \citep{rw10}. The two pole caustic (TPC) model has great difficulty
producing the $\gamma$-ray pulse width $\Delta = 0.23$ and 
lag from the radio $\delta=0.29$. In the outer gap (OG) picture, $\gamma$-rays
for such an old pulsar are only seen if viewed from near the equator,
$\zeta > 60^\circ$, although the observed pulse width and $\gamma$-radio lag
are then easily achieved. The best-fit prefers $\zeta \approx
65-75^\circ$, consistent with the angles inferred from the bow shock
geometry above. 

\section{Conclusions}

	The discovery of an H$\alpha$ bow shock associated with the $\gamma$-ray pulsar
PSR J1741$-$2054 provides an important opportunity to constrain the geometry
and momentum deposition of a pulsar wind.  Our initial images and spectra 
suggest that this pulsar wind has a strong equatorial concentration and that 
the spin axis (and space velocity) are close to the plane of the sky. 
The spectroscopy implies a low extinction (close distance) for the nebula,
consistent with pulsar dispersion measure estimates.
Spectra were obtained for two slices across the nebula which are best described
by a pulsar velocity of $\sim 150{\rm \, km\,s}^{-1}$ 
directed $15 \pm 10^\circ$ out of the plane of the sky.
The variation in the spectrum along the nebula's velocity
axis suggests that different layers (prompt and then mixed) of the post-shock 
flow are emitting near the nebula apex and tail respectively. Some unmodeled features are seen
and numerical simulation including pressure effects in the shock tail may be
needed to explain the full velocity structure.

The apparently equatorially concentrated
wind and near edge-on view are consistent with inferences of PWN flow
from observations of other pulsars and with the $\gamma$-ray emission geometry
inferred from LAT pulsations. 

	Pulsar H$\alpha$ bow shocks often display high symmetry suggesting that
careful modeling can exploit their spatial and velocity structure to probe
the dynamics of collisionless shocks in partly ionized media at a range
of inclination angles. If, as suggested by our initial imaging and spectroscopy,
the portion of the post shock flow probed by the H$\alpha$ emission varies 
along the bow shock surface, then the shock of PSR J1741$-$2054 provides
an especially important laboratory to study the structure of
collisionless non-radiative shocks. Improved measurements could precisely determine
the anisotropy of the relativistic pulsar wind, may constrain
the degree of wind-velocity alignment, and can help deliver a kinematic
measurement of the pulsar distance. Such studies will, however, 
require higher spatial and spectral resolution data at good sensitivity.  This 
might be acquired through intermediate resolution integral field studies
or Fabry-Perot H$\alpha$ scans \citep{mrf99}.

\bigskip

	We thank Tony Readhead for collaboration on {\it Fermi} counterpart studies 
and for the LRIS integrations, and Aaron Meisner for help with the WIYN observing 
and MiniMo processing. We also thank Pablo Saz-Parkinson,
Scott Ransom and Paul Ray for valuable discussions on the detection and distance of the
LAT pulsars. Finally, we thank the referee, Marten van Kerkwijk, for a prompt, 
careful reading and useful suggestions that improved the text.
This work was supported in part by NASA grants NNX08AW30G and NNX10AD11G and
Chandra Grant GO0-11097X.

\end{document}